\documentclass[pra,aps,amssymb,amsmath,amsmath,showpacs,reprint,twocolumn]{revtex4-1}
\usepackage{color}
\usepackage{graphicx}
\usepackage{epstopdf}
\usepackage{natbib}
\usepackage{hyperref}
\usepackage{dcolumn}
\usepackage{bm}
\usepackage{dcolumn}
\usepackage{multirow}

\newcolumntype{d}[1]{D{.}{.}{#1}}

\newcommand{\Fkt}[1]{\,\mathsf {#1}}
\def\openone{\leavevmode\hbox{\small1\kern-3.3pt\normalsize1}}

\ifx\Tr\renewcommand{\Tr}{\Fkt{Tr}} %has to commented out for IOP
\else\newcommand{\Tr}{\Fkt{Tr}}
\fi

\newcommand{\vecr}{\mathbf{r}}

\usepackage{booktabs}
\usepackage{graphicx}
\usepackage{amsmath}
\usepackage{indentfirst}
\begin{document}
\title{Atomic Bethe logarithm in the mean-field approximation}

\author{\sc Micha\l\ Lesiuk}
\email{e-mail: m.lesiuk@uw.edu.pl}
\author{\sc Jakub Lang}
\affiliation{\sl Faculty of Chemistry, University of Warsaw\\
Pasteura 1, 02-093 Warsaw, Poland}
\date{\today}

\begin{abstract}
In this work we develop and implement a method for calculation of the Bethe logarithm for many-electron atoms. This quantity is required to evaluate the leading-order quantum electrodynamics correction to the energy and properties of atomic and molecular systems beyond the Dirac theory (the Lamb shift). The proposed formalism is based on the mean-field representation of the ground-state electronic wavefunction and of the response functions required in the Schwartz method [C. Schwartz, Phys. Rev. {\bf 123}, 1700 (1961)]. We discuss difficulties encountered in the calculations with the emphasis on the specific basis set requirements in the vicinity of the atomic nucleus. This problem is circumvented by introducing a modified basis set of exponential functions which are able to accurately represent the gradient of hydrogen-like orbitals. The Bethe logarithm is computed for ground electronic states of atoms from hydrogen to magnesium and, additionally, for argon. Whenever possible, the results are compared with the available reference data from the literature. In general, the mean-field approximation introduces a surprisingly small error in the calculated values, suggesting that the electron correlation effects are of minor importance in determination of the Bethe logarithm. Finally, we propose a robust scheme to evaluate the Lamb shift for arbitrary light molecular systems at little computational cost. As an illustration, the method is used to calculate Lamb shifts of the vibrational levels of the nitrogen molecule.
\end{abstract}

\maketitle

\section{Introduction}
\label{sec:intro}

In recent years, there has been growing interest in determination of the effects of quantum electrodynamics (QED) on the structure and properties of many-electron atoms and molecules. While the theory of bound-state QED was formulated decades ago and has been successfully used since then, it was believed that the QED corrections are negligible for such systems. However, as the accuracy of theoretical calculations improves, the necessity to account for the QED effects becomes apparent in at least two situations. First, in systems involving heavy atoms the QED effects become large enough to constitute the major source of the overall error. This is best illustrated by recent calculations of the ionization energy and electron affinity of the gold atom by Pa\v{s}teka and collaborators~\cite{pasteka17}. Only after the QED effects were taken into consideration, it was possible to bring the theory and experiment to an agreement of around $1\,$meV. The second reason for the growing interest in the QED effects is the demand for highly accurate calculations for many-electron atoms and small molecules. For example, recent calculations for hydrogen molecule~\cite{piszczatowski09}, helium dimer~\cite{cencek12,czachorowski20}, neon~\cite{lesiuk20b,hellmann22} and argon~\cite{lesiuk23} atoms, berylium dimer~\cite{lesiuk15,lesiuk19}, NaLi molecule~\cite{gronowski20}, lithium dimer~\cite{lesiuk20}, and He$_2^+$ molecular ion~\cite{gebala23} had to include the QED corrections to reach the required accuracy, despite the fact that these systems do not involve heavy atoms and hence these corrections are small by the usual standards.

In this work we focus on the leading-order QED corrections (the Lamb shift) which take into account the effects of vacuum polarization and electron self-energy. Determination of these corrections for arbitrary atomic and molecular systems is a non-trivial task and several approaches are available in the literature for this purpose. This includes Uehling~\cite{uehling35} and Wichmann-Kroll potentials for vacuum polarization~\cite{wichmann56,blomqvist72}, local effective potentials inspired by the hydrogen Lamb shift proposed by Pyykk\"{o} and collaborators~\cite{pyykko98,labzowsky99b,pyykko01,dyall01,pyykko03}, model operator approach of Shabaev et al.~\cite{shabaev13,shabaev22}, multiple commutator approach proposed by Labzowsky and Goidenko~\cite{labzowsky97,labzowsky99}, and effective Hamiltonians of Flambaum and Ginges~\cite{flambaum05}. However, this list is not exhaustive. Some of these methods are already available in mainstream atomic electronic structure programs~\cite{dyall89,shabaev15,kahl19}. For molecules, the studies including the QED effects are less numerous. They are based primarily on effective local potentials fitted by a linear combination of Gaussian functions, see the papers of Pyykkö and Zhao~\cite{pyykko03} and  Peterson and collaborators~\cite{shepler05,shepler07,peterson15}, as well as references therein. The QED effects were also included in several core pseudopotentials for heavy elements designed by the Stuttgart/Cologne group~\cite{hangele12,hangele13a,hangele13b,hangele14}. Finally, in the recent work by Sunaga et al.~\cite{sunaga22}, the Flambaum-Ginges effective potential has been adopted to molecular calculations and implemented in Dirac code~\cite{dirac23}. This paper includes also an excellent introduction to the bound-state QED formalism accessible to non-experts.

Our interest in the QED effects is driven primarily by the applications in metrology~\cite{gaiser15,rourke19,gaiser20,gaiser22}, where theoretical predictions for systems such as light noble gases or major constituents of Earth's atmosphere are important ingredients. In particular, to calculate the quantities such as polarizability, magnetic susceptibility, dielectric and pressure virial coefficients of atomic and molecular gases with accuracy of one part per thousand (or better), the QED effects have to be routinely taken into account. As the systems relevant in the aforementioned applications are composed of light atoms (the first and second row of the periodic table), both the relativistic and QED corrections can be calculated perturbatively. This greatly simplifies the formalism as the non-relativistic two-component wavefunction, which is the basis of the perturbative approach, can be calculated using well-developed tools.

For light atoms, probably the most consistent way to account for the QED effects is offered by the non-relativistic QED theory (NRQED) introduced by Caswell and Lepage~\cite{caswell86}, and further developed and extended by Pachucki~\cite{pachucki93,pachucki98,pachucki04,pachucki05}. 
This formalism is based on expansion of the energy (and other properties) in the powers of the fine structure constant ($\alpha\approx 137.036$)
\begin{align}
 E = E_0 + \alpha^2 E_{\mathrm{rel}} + \alpha^3 E_{\mathrm{QED}} + \ldots
\end{align}
If the Born-Oppenheimer approximation is assumed, the first term ($E_0$) is the non-relativistic energy corresponding to the clamped-nuclei Schr\"{o}dinger-Coulomb Hamiltonian. The relativistic correction of the order $\alpha^2$, denoted as $E_{\mathrm{rel}}$, is defined the expectation value of the Breit-Pauli Hamiltonian~\cite{bethe57}. Nowadays it can be routinely evaluated, at least for closed-shell systems~\cite{coriani04}. The leading-order pure QED correction (not included in the Dirac-Coulomb equation) reads~\cite{araki57,sucher58}
\begin{align}
\label{qed}
\begin{split}
E_{\mathrm{QED}} &= \frac{8\alpha}{3\pi}\left(\frac{19}{30}-2\ln \alpha - \ln k_0\right)\langle D_1 
\rangle \\
&+ \frac{\alpha}{\pi} \left(\frac{164}{15}+\frac{14}{3}\ln \alpha\right)\langle D_2 \rangle + 
\langle
H_{AS} \rangle,
\end{split}
\end{align}
where $\ln k_0$ is the Bethe logarithm \cite{bethe57} defined precisely further in the text,
$\langle D_1 \rangle$ and $\langle D_2 \rangle$ are the one- and two-electron Darwin terms 
\begin{align}
\langle D_1 \rangle &= \frac{\pi}{2}\,\alpha^2\sum_A Z_A \langle \sum_n 
\delta(\textbf{r}_{nA})\rangle,\\
\langle D_2 \rangle &= \pi\,\alpha^2\,\langle\sum_{n>n'}\delta(\textbf{r}_{nn'})\rangle,
\end{align}
where $Z_A$ are the nuclear charges and $\delta (\vecr)$ is the three-dimensional Dirac delta distribution. Throughout the paper, we use the indices $n,n',\ldots$ and $A,B,\ldots$ to denote electrons and nuclei, respectively. A shorthand notation $\langle X\rangle$ is employed for the expectation value of an operator $X$ on the ground state wavefunction $\Psi_0$, i.e., $\langle X\rangle = \langle \Psi_0 | X \Psi_0 \rangle$. The last term in Eq. (\ref{qed}) is the Araki-Sucher correction $\langle H_{AS} \rangle$, see Eqs. (4)--(6) in Ref.~\cite{balcerzak17} for a precise definition in a notation consistent with the present work. 

In this work we consider evaluation of the leading-order QED correction ($E_{\mbox{\scriptsize 
QED}}$) for light many-electron systems, both atomic and molecular, in the ground electronic state. In this context, two quantities appearing in Eq. (\ref{qed}) are non-trivial to evaluate, namely $\langle H_{AS} \rangle$ and $\ln k_0$. Concerning the former, the results given in Ref.~\cite{balcerzak17} suggest that large Gaussian basis sets combined with a theoretically-motivated extrapolation formula offer a sufficient level of accuracy for most applications to many-electron systems. Therefore, the major difficulty now lies in computation of the Bethe logarithm, $\ln k_0$. It is formally defined by the expression
\begin{align}
\label{bethe0}
\ln k_0 = \frac{\mathcal{N}}{\mathcal{D}} = \frac{\langle
\Psi_0|\nabla(H-E_0)\ln 2|H-E_0|\nabla|\Psi_0\rangle}{\langle 
\Psi_0|\nabla(H-E_0)\,\nabla|\Psi_0\rangle},
\end{align}
where $\hat{H}$ is the non-relativistic Hamiltonian of the system and $\Psi_0$ is the corresponding 
wavefunction, and $\nabla=\sum_n \nabla_n$ is the (total) electronic momentum operator. 

There are two main approaches available in the literature for calculation of the Bethe logarithm. The first one, due to Goldman and Drake~\cite{drake00}, uses the closure relation to rewrite Eq. (\ref{bethe0}) as an infinite summation over excited states of the system (or a set of pseudostates formed in a given basis). As this sum is slowly convergent, a large number of excited state energies and wavefunctions need to be found. This limits the applicability of the Drake-Goldman method to, mostly, one- and two-electron atoms and molecules (see Refs.~\cite{drake88,drake90,goldman94,bhatia98,korobov04,zhong13} and references therein), and computations for larger systems that employ this formalism are scarce~\cite{yan03,yan08}.

The second method was put forward by Schwartz in his seminal work concerning the helium 
atom~\cite{schwartz61}. The Schwartz method involves a representation of the Bethe logarithm as an integral over virtual photon momenta. This leads to a formula which is much more manageable from the point of view of applications to many-electron systems. Thus far, this technique (with subsequent modifications) has been applied to helium~\cite{schwartz61,korobov99,baker00,korobov12,korobov19}, 
lithium~\cite{yan98,pachucki03,yan03,puchalski13} and beryllium atoms~\cite{pachucki04b,puchalski13b}, and some of the corresponding ions~\cite{pachucki06,bubin10}, and to a handful of molecular systems, namely H$_2^+$~\cite{bukowski92,korobov04b,korobov06,korobov12,korobov12b}, H$_2$~\cite{kpisz09,komasa11,saly23}, He$_2^+$ cation and H$_3$~\cite{ferenc23}.

The aforementioned computations of the Bethe logarithm for few-electron systems were based on Hylleraas-like~\cite{hyll29} or explicitly correlated Gaussians (ECG) basis sets~\cite{mitroy13} for expansion of the electronic wavefunction and other necessary quantities. Clearly, neither approach can easily be extended to systems with more than several electrons. In this work we make the first step towards introduction of general techniques to calculate $\ln k_0$ from wavefunctions expanded in a set of one-electron orbitals. The main advantage of this approach is that it can be applied, in principle, to any atomic and molecular system for which the perturbative QED expansion of the energy makes sense.

In this work we adopt the Hartree-Fock method to model the electronic wavefunction, $\Psi_0$. While this approach is not accurate enough for some applications, it is a necessary stepping stone for more rigorous treatments involving, e.g. coupled-cluster wavefunctions~\cite{crawford07,bartlett07}. We show that this approximation leads to surprisingly accurate results, suggesting that the electron correlation effects are of secondary importance in calculation of $\ln k_0$. Importantly, we present a novel method to avoid basis set incompleteness problems in the Schwartz method which is the most serious obstacle in calculation of the Bethe logarithm with one-electron basis sets. It is shown that the basis set of Slater-type orbitals~\cite{slater30,slater32} with inclusion of non-standard $1p$, $2d$, $\ldots$, functions provide suitable representation of all auxiliary quantities appearing in the computations. In this paper we focus on the Bethe logarithm for one-center systems (neutral atoms and their cations/anions), but discuss approximations suitable for arbitrary molecular systems.

The atomic units are used throughout the present work unless explicitly stated otherwise. The Einstein convention (summation over repeating indices) is assumed in all expressions. We assume that all orbitals and wavefunctions are purely real.

\section{Molecular Bethe logarithm}
\label{sec:molbethe}

As mentioned in the introduction, in this work we focus on calculations of the Bethe logarithm for one-center systems. On the surface, this appears to have a limited usefulness in the applications discussed above. However, knowing the value of the Bethe logarithm for isolated atoms, one can easily generate a reasonable approximation to $E_{\mathrm{QED}}$ for arbitrary molecules composed of these atoms. This follows from the weak dependence of the Bethe logarithm on the molecular geometry. While this fact has been noted in the literature, and has been used in practice~\cite{lesiuk15,lesiuk19,lesiuk20,gronowski20,gebala23}, we
believe that a comprehensive discussion of this phenomena is in order. Therefore, in this section we provide general expressions for the approximate molecular $E_{\mathrm{QED}}$ and gather the evidence available in the literature that supports it.

Let us consider a diatomic molecule $\mbox{A}\mbox{B}$ composed of the atoms $\mbox{A}$, $\mbox{B}$. Generalization of this formalism to polyatomic molecules is straightforward. The Bethe logarithm for $\mbox{A}\mbox{B}$ is denoted by $\ln k_0(\mbox{A}\mbox{B})$ and similarly $\ln k_0(\mbox{X})$, $\mbox{X}=\mbox{A},\mbox{B}$, stand for the Bethe logarithms of the isolated subsystems. An analogous notation $\langle D_1\rangle_{\mathrm{AB}}$, $\langle D_1\rangle_{\mathrm{A}}$, $\langle D_1 \rangle_{\mathrm{B}}$ is preserved for the one-electron Darwin corrections of all species involved. Assuming that $\ln k_0(\mbox{A}\mbox{B})$ is weakly dependent on the molecular geometry, we can replace it by a value obtained for an arbitrary chosen interatomic distance, $R$. It is convenient to select as the reference geometry the asymptotic value of $R$, corresponding to the non-interacting limit $R\rightarrow\infty$. Therefore, it remains to derive the proper formula for $\ln k_0(\mbox{A}\mbox{B})$ valid for $R\rightarrow\infty$. To this end, we take advantage of the fact that 
the total QED energy correction, Eq. (\ref{qed}), must be extensive. In other words, the value of $E_{\mathrm{QED}}$ for $\mbox{A}\mbox{B}$ is equal to the sum of corrections for the isolated subsystems as $R\rightarrow\infty$. By equating the expressions for $E_{\mathrm{QED}}$ for the diatom and the sum of $E_{\mathrm{QED}}$ for both atoms,
and eliminating all quantities that are manifestly extensive we arrive at the 
following relation valid in the non-interacting limit
\begin{align}
\begin{split}
 \ln k_0(\mbox{A}\mbox{B})\cdot\langle D_1\rangle_{\mathrm{AB}} &= \ln k_0(\mbox{A}) 
\cdot\langle D_1\rangle_{\mathrm{A}} \\
&+\, \ln k_0(\mbox{B})\cdot\langle D_1\rangle_{\mathrm{B}},
\end{split}
\end{align}
and since the Darwin correction is extensive, i.e., $\langle 
D_1\rangle_{\mathrm{AB}}=\langle D_1\rangle_{\mathrm{A}}+\langle D_1 
\rangle_{\mathrm{B}}$ as $R\rightarrow\infty$, one finally has
\begin{align}
\label{apprmol}
 \ln k_0(\mbox{A}\mbox{B}) = \frac{\ln k_0(\mbox{A}) 
\cdot\langle D_1\rangle_{\mathrm{A}} + \ln 
k_0(\mbox{B})\cdot\langle D_1\rangle_{\mathrm{B}}}{\langle 
D_1\rangle_{\mathrm{A}}+\langle D_1 
\rangle_{\mathrm{B}}},
\end{align}
where the right-hand-side depends only on the properties of the isolated atoms. This is the approximate formula for the molecular Bethe logarithm we recommend for arbitrary geometries.

\begin{figure}[t!]
 \includegraphics[scale=0.58]{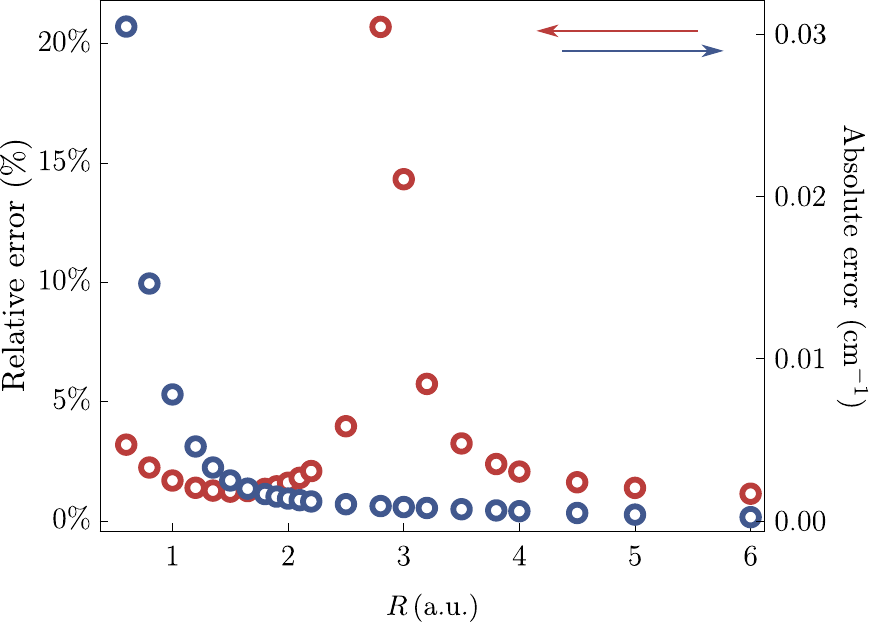}
 \caption{Relative errors in (in \%, red dots) and absolute errors (in cm$^{-1}$, blue dots) in the QED correction for the H$_2^+$ molecular ion as a function of the internuclear distance, $R$. Data taken from Ref.~\cite{bukowski92}.\label{fig:h2plus}}
\end{figure}

\begin{figure}[t!]
 \includegraphics[scale=0.58]{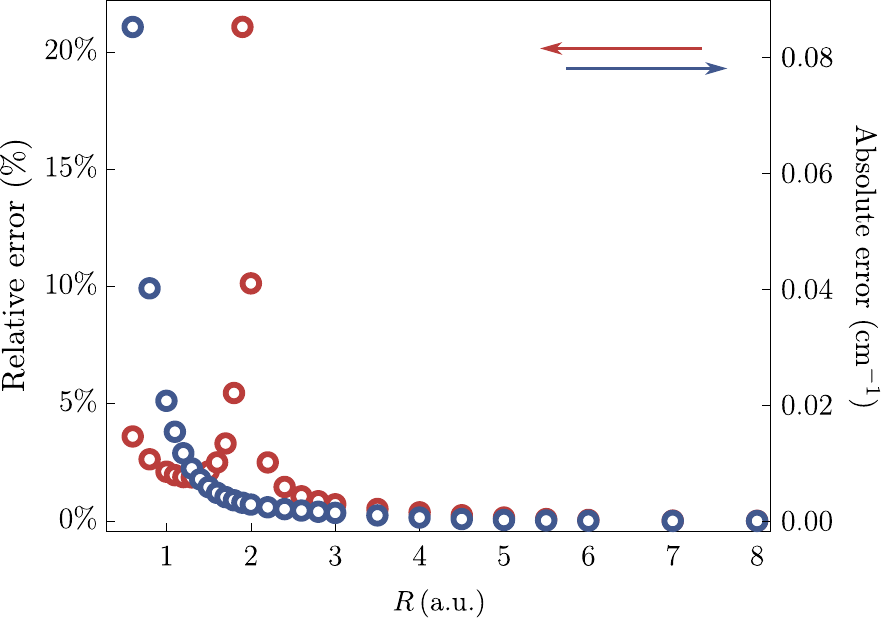}
 \caption{Relative errors in (in \%, red dots) and absolute errors (in cm$^{-1}$, blue dots) in the QED correction for the H$_2$ molecule as a function of the internuclear distance, $R$. Data taken from Ref.~\cite{kpisz09}.\label{fig:h2}}
\end{figure}

To study the accuracy of this approximation, we analyze the data available in the literature  and compare with the molecular $\ln k_0$ calculated explicitly for a set of internuclear distances with a controlled precision. We consider the H$_2^+$ molecular ion and H$_2$ molecule within the Born-Oppenheimer approximation. The Bethe logarithms for these systems were calculated by Bukowski et al.~\cite{bukowski92} and Piszczatowski et al.~\cite{kpisz09}, respectively. In Fig.~\ref{fig:h2plus} and Fig.~\ref{fig:h2} we provide a comparison of the reference values of the Bethe logarithm for this systems with the approximation advocated in this work, Eq.~(\ref{apprmol}). For both molecules, the absolute errors in the QED correction determined according to Eq.~(\ref{apprmol}) is, on average, below $0.01\,$cm$^{-1}$. In the plot of relative errors, we see a sharp spike at $R\approx 3\,$a.u. for H$_2^+$ and $R\approx 2\,$a.u., where they reach around 20\%. However, this feature is due to the fact that the QED correction crosses zero and changes sign. Therefore, even a large relative error is this regime does not translate into a significant absolute error. On average, the relative errors are of the order of 2-3\% for both molecules.

\section{Theory}
\label{sec:theory}

\subsection{Wavefunction model}
\label{sec:wfn}

Consider the calculation of the Bethe logarithm for first- and second-row atoms. The ground-state electronic states of these systems are characterized by non-relativistic terms $^1$S, $^2$S, $^2$P, $^3$P, and $^4$S. In general, the electronic wavefunction of these atoms cannot be represented as a single Slater determinant without violating the spin and spatial symmetry of the system. While this deficiency of the single-determinantial approach is often neglected in favor of simplicity, we adopt a more general open-shell Hartee-Fock method which avoids this problem. Following Roothaan~\cite{roothaan60}, the wavefunction is represented as a sum of several Slater determinants, each containing the common set of doubly-occupied closed-shell orbitals and a fixed number of partially-occupied open orbitals chosen from a predefined set. In the following, closed-shell orbitals are denoted by the indices $i,j,\ldots$, open orbitals by $x,y,\ldots$, and virtual (empty) orbitals by $a,b,\ldots$. The indices $p,q,\ldots$ are used for the whole orbital space, i.e. $\{ p \} = \{i\}\cup\{x\}\cup\{a\}$.
Each determinant entering the wavefunction has an equal weight, fixed by the normalization condition of the state. The total energy of the atom is given by
\begin{align}
\label{erohf}
\begin{split}
    E &= 2\sum_i (i|h|i) + \sum_{ij}\Big[ 2(ii|jj) - (ij|ji) \Big] \\
    &+ 2f\bigg[ \sum_x (x|h|x) + \sum_{ix} \Big[ 2(xx|ii) - (xi|ix) \Big] \bigg] \\
    &+ f^2\sum_{xy} \Big[ 2a(xx|yy) - b (xy|yx) \Big],
\end{split}
\end{align}
where $h$ is the one-electron Hamiltonian, and $(pq|rs)$ are the two-electron integrals in the Coulomb notation. The parameters $f$, $a$, and $b$ are dependent on the electronic configuration; for all states considered in this work their values are given in Ref.~\cite{roothaan60}.

The orbitals from which the determinants are constructed are found by employing the variational principle. Minimizing the expression~(\ref{erohf}) with respect to the orbitals (with necessary orthonormality constraints) leads to a set of single-particle equations that can be solved self-consistently. The orbitals are found as eigenfunctions of an effective Fock operator, $R$. This operator is not defined uniquely, see the discussion in Refs.~\cite{plakhutin14a,plakhutin14b}, and in this work we follow the parametrization of Hirao and Nakatsuji~\cite{hirao73}:
\begin{align}
\label{rcoupling}
    R =
    \begin{bmatrix}
        F_C & F_O - F_C & F_C \\
        F_O - F_C & F_O & F_O \\
        F_C & F_O & F_C + F_O \\
    \end{bmatrix},
\end{align}
where the three rows/columns of this matrix correspond to the closed, open, and virtual (unoccupied) orbital manifolds. The Fock operators of closed and open manifolds, $F_C$ and $F_O$, respectively, are defined as
\begin{align}
    F_C &= h + \big[ 2 J_C - K_C \big] + f\big[ 2 J_O - K_O \big], \\
    F_O &= f \Big[ h + \big[ 2 J_C - K_C \big] + f\big[ 2a J_O - b K_O \big] \Big],
\end{align}
where the Coulomb and exchange operators acting on an arbitrary one-electron function $\phi(\mathbf{r})$ give
\begin{align}
    J_C\phi(\mathbf{r}) = \phi(\mathbf{r})\sum_i\int d\mathbf{r}'\;
    \frac{\varphi_i^*(\mathbf{r}')\,\varphi_i(\mathbf{r}')}{|\mathbf{r}-\mathbf{r}'|}, \\
    K_C\phi(\mathbf{r}) = \sum_i \varphi_i(\mathbf{r})\int d\mathbf{r}'\;
    \frac{\varphi_i^*(\mathbf{r}')\,\phi(\mathbf{r}')}{|\mathbf{r}-\mathbf{r}'|},
\end{align}
and analogously for the open manifold. At convergence of the self-consistent field procedure, the operator $R$ diagonal in the whole orbital basis. The eigenvalues (the orbital energies) are denoted by the symbols $\epsilon_i$, $\epsilon_x$, and $\epsilon_a$ for the closed, open, and virtual orbital manifolds, respectively.

\subsection{The Schwartz method with approximate wavefunctions}
\label{sec:schwartz}

The Schwartz method relies on the following operator relation valid in the $K\rightarrow\infty$ limit 
\begin{align}
\label{schwint}
 \ln 2|H-E_0| \approx \ln 2K - \int_0^K \frac{dk}{H-E_0+k},
\end{align}
which is inserted into the numerator of Eq.~(\ref{bethe0}). Note that the both terms on the 
right-hand-side of Eq.~(\ref{bethe0}) diverge for large $K$ and this divergence must be canceled out numerically. The equation~(\ref{schwint}) provides the basis for various integral representations of the Bethe logarithm. A particularly convenient and compact formula was given by Pachucki and Komasa~\cite{pachucki04b}
\begin{align}
\label{lnk0_t}
\begin{split}
 &\ln k_0 = -\frac{1}{\mathcal{D}}\int_0^1 dt\;\mathcal{F}(t), \\
 &\mathcal{F}(t) = \frac{f(t)+\langle\Psi_0|\nabla^2|\Psi_0\rangle+2\mathcal{D}t^2}{t^3},
\end{split}
\end{align}
where
\begin{align}
\label{ft}
 f(t) = \big\langle\Psi_0\big| \nabla \frac{k}{H-E_0+k} \nabla \big|\Psi_0\big\rangle,
\end{align}
and the variables $k$ and $t$ are related through the relation $t=(1+2k)^{-1/2}$. Note that a 
different sign convention is used in Eqs.~(\ref{lnk0_t})~and~(\ref{ft}) as compared with 
Ref.~\cite{pachucki04b}.

Provided that the electronic wavefunction used in Eq.~(\ref{bethe0}) is exact, we
have the following expression for the denominator
\begin{align}
\label{exd}
 \mathcal{D} = -2\pi\sum_A Z_A \langle \sum_n \delta(\textbf{r}_{nA})\rangle.
\end{align}
Unfortunately, this expression is not valid for approximate wavefunction models which do not 
satisfy the electronic Schr\"{o}dinger equation, $(H-E_0)\Psi_0=0$. In particular, this is true for the Hartree-Fock wavefunction employed in this work to approximate $\Psi_0$. Following the usual practice, one could evaluate $\mathcal{D}$ by simply inserting an approximate wavefunction into Eq.~(\ref{exd}). Unfortunately, this choice would make the integrand in Eq.~(\ref{lnk0_t}) singular for small $t$ because the terms proportional to $t^2$ in $f(t)$ would no longer be canceled out by the  $2\mathcal{D}t^2$ counterterm. This may lead to instabilities or even divergence of the numerical integration in Eq.~(\ref{lnk0_t}). To eliminate this problem, we propose a different method of evaluating the denominator, described in detail in Sec.~\ref{sec:denominator}.

The small-$t$ expansion of the function $f(t)$ reads~\cite{schwartz61,bukowski92}
\begin{align}
\label{ftsml}
 f(t) = -\langle\nabla^2\rangle - 2\mathcal{D}t^2 + f_3 t^3 + f_4 t^4 \ln t + \mathcal{O}(t^4),
\end{align}
which is valid (in the complete basis set limit) both for exact and approximate wavefunctions 
provided that the denominator is calculated according procedure described in Sec.~\ref{sec:denominator}. This expansion is used to avoid direct computation of $f(t)$ for very small $t$ which may be unstable.
In the case of the exact wavefunction explicit expressions are known also for the higher-order 
coefficients in this expansion ($f_3$, $f_4$)~\cite{forrey93}. We attempted to derive the 
corresponding expressions for approximate wavefunctions along the lines of Ref.~\cite{forrey93}. Unfortunately, the final formulas were not tractable for systems larger than two electrons, so that the higher-order coefficients must be obtained by fitting. Details of this procedure are given further in the text.

\subsection{Response-like equations}
\label{sec:response}

Provided that the value of $t$ is not too small the quantity $f(t)$ is accurately calculated from the 
following response-like equation
\begin{align}
\label{psik}
 \left(H-E_0+k\right) \Psi(k) = \nabla \Psi_0,
\end{align}
so that $f(t) = k\,\langle \nabla \Psi_0 | \Psi(k) \rangle$, cf. Eq. (\ref{ft}). Note that as the values of $k$ required in Eq. (\ref{lnk0_t}) are real and non-negative the calculations are always performed away from the poles of the resolvent, $\left(H-E_0+k\right)^{-1}$, and the 
operator $H-E_0+k$ is Hermitian.

For the sake of convenience, we expand the wavefunction $\Psi(k)$ into a linear combination of excited-state determinants. By employing Slater-Condon rules it is straightforward to show that only singly-excited determinants bring a non-zero contribution to Eq.~(\ref{psik}). Therefore, we first generate a complete set of singly-excited determinants, i.e. obtained from the reference determinants present in $\Psi_0$ by replacement of one orbital. In general, there are three classes of such determinants obtained by:
\begin{itemize}
 \item replacing an open orbital $x$ by a virtual orbital $a$;
 \item replacing a closed orbital $i$ by an open orbital $x$
 (the latter must not present in the determinant already);
 \item replacing a closed orbital $i$ by a virtual orbital $a$;
\end{itemize}
We denote the expansion coefficients corresponding to these classes by $U_{ax}$, $U_{xi}$, and $U_{ai}$, respectively. By inserting this expansion in place of $\Psi(k)$ in Eq.~(\ref{psik}) and projecting the resulting equation onto all singly-excited determinants, one obtains three sets of linear equations:
\begin{widetext}
\begin{align}
\begin{split}
    &(x|G_O-G_C|i) + \sum_{j} U_{xj}(j|F_O|i) + \sum_{a} U_{ax}(a|F_O|i)
    - \sum_{a} U_{ai}(x|F_C|a) \\
    &- \sum_{y} U_{yi}(x|F_C|y) - U_{xi}\big[\epsilon_x + \epsilon_i - k( f - 1 ) \big] = (1-f)(x|\nabla|i) \\
    &(a|G_C|i) - \sum_{b} U_{bi} (a| F_O |b) 
    - \sum_{x} U_{ax} (x| F_C |i) - \sum_{x} U_{xi} (a| F_C |x) + U_{ai}
    \big( \epsilon_a - \epsilon_i + k\big) = -(a|\nabla|i) \\
    &(a|G_O|x) - \sum_{i} U_{ai} (i|F_O |x) +\sum_{i} U_{xi}(a|F_O|i) + 
    U_{ax} \big( \epsilon_a - \epsilon_x + kf \big) = -f(a|\nabla|x),
\end{split}
\end{align} 
\end{widetext}
where $G_O$ and $G_C$ are the perturbed Fock operators in the open and closed orbital manifold, respectively, given by
\begin{align}
\label{goc1}
\begin{split}
    G_C &= \big[ 2 J_C' - K_C' \big] + f\big[ 2 J_O' - K_O' \big], \\
    G_O &= \big[ 2 J_C' - K_C' \big] + f\big[ 2a J_O' - b K_O' \big] \Big],
\end{split}
\end{align}
and the perturbed Coulomb and exchange operators are defined through their action on an arbitrary orbital $\phi(\mathbf{r})$ as follows 
\begin{align}
\label{goc2}
\begin{split}
    J_C'\phi(\mathbf{r}) &=
    2\phi(\mathbf{r})\sum_{bj} U_{bj} \int d\mathbf{r}'\;\frac{\varphi_b^*(\mathbf{r}')\,\varphi_j(\mathbf{r}')}{|\mathbf{r}-\mathbf{r}'|} \\
    &- 2\phi(\mathbf{r})\sum_{yj} U_{yj} \int d\mathbf{r}'\;\frac{\varphi_y^*(\mathbf{r}')\,\varphi_j(\mathbf{r}')}{|\mathbf{r}-\mathbf{r}'|},
\end{split}
\end{align}
\begin{align}
\label{goc3}
\begin{split}
    &K_C'\phi(\mathbf{r}) =
    \sum_{bj} U_{bj} \bigg[ \varphi_j(\mathbf{r})
    \int d\mathbf{r}'\;\frac{\varphi_b^*(\mathbf{r}')\,\phi(\mathbf{r}')}
    {|\mathbf{r}-\mathbf{r}'|} \\
    &+ \varphi_b(\mathbf{r})
    \int d\mathbf{r}'\;\frac{\varphi_j^*(\mathbf{r}')\,\phi(\mathbf{r}')}
    {|\mathbf{r}-\mathbf{r}'|} \bigg] \\
    &- \sum_{yj} U_{yj} \bigg[ \varphi_j(\mathbf{r})
    \int d\mathbf{r}'\;\frac{\varphi_y^*(\mathbf{r}')\,\phi(\mathbf{r}')}
    {|\mathbf{r}-\mathbf{r}'|} \\ 
    &+ \varphi_y(\mathbf{r})
    \int d\mathbf{r}'\;\frac{\varphi_j^*(\mathbf{r}')\,\phi(\mathbf{r}')}
    {|\mathbf{r}-\mathbf{r}'|} \bigg],
\end{split}
\end{align}
and analogously for the open manifold. Due to large size, the above equations are solved iteratively. In our implementation we use the preconditioned conjugate gradient method. Once the response equations are solved, the main quantity of interest, namely $f(t)$ defined by Eq.~(\ref{ft}), is straightforward to calculate
\begin{align}
\begin{split}
    &f(t) = 4k(f-1)\sum_{xi} U_{xi} (x|\nabla|i) \\
    &+ 4k\sum_{ai} U_{ai} (a|\nabla|i) + 4fk \sum_{xa} U_{ax} (a|\nabla|x).
\end{split}
\end{align}

\subsection{Treatment of the denominator}
\label{sec:denominator}

In this section, we propose an alternative method to evaluate the denominator in Eq.~(\ref{bethe0}).
In this approach, the requirement that the Schr\"{o}dinger equation is solved exactly is replaced by a much weaker condition that the one-electron basis used for expansion of the Hartree-Fock orbitals is complete. Under this assumption, the gradient of the closed and open orbitals can be expanded in the orbital basis as follows
\begin{align}
 \nabla |i) = \sum_p |p)(p|\nabla|i), \;\;\;
 \nabla |x) = \sum_p |p)(p|\nabla|x).
\end{align}
The gradient of the Hartree-Fock wavefunction $\Psi_0$ becomes a linear combination of singly-excited determinants. Derivation of the explicit expression for the denominator is hence accomplished using the standard Slater-Condon rules, giving
\begin{align}
\label{newden}
\begin{split}
 \mathcal{D} = &-6\sum_{ix} R_{ix} (i|\nabla|x) 
 -6\sum_{ai} R_{ai} (a|\nabla|i) \\
 &-6\sum_{ax} R_{ax} (a|\nabla|x),
\end{split}
\end{align}
where
\begin{widetext}
\begin{align}
\begin{split}
    R_{ix} &= (x|G_O-G_C|i) + \sum_{j} (j|\nabla|x) (j|F_O|i) + \sum_{a} (a|\nabla|x) (a|F_O|i) \\
    &- \sum_{a} (a|\nabla|i) (x|F_C|a) + \sum_{y} (i|\nabla|y) (x|F_C|y) - 
    (i|\nabla|x) \big(\epsilon_x + \epsilon_i \big), \\
    R_{ai} &= (a|G_C|i) - \sum_{b} (b|\nabla|i)  (a| F_O |b) - \sum_{x} (a|\nabla|x) (x| F_C |i) - \sum_{x} (i|\nabla|x)  (a| F_C |x) + (a|\nabla|i) \big( \epsilon_a - \epsilon_i \big), \\
    R_{ax} &= (a|G_O|x) - \sum_{i} (a|\nabla|i)  (i|F_O |x) +\sum_{i} (i|\nabla|x) (a|F_O|i) + 
    (a|\nabla|x)  \big( \epsilon_a - \epsilon_x \big).
\end{split}
\end{align} 
\end{widetext}
In the above formulas, the operators $G_C$ and $G_O$ are defined in exactly same way as in Eqs.~(\ref{goc1})-(\ref{goc3}) with the exception that $U_{ax}$, $U_{ai}$, and $U_{xi}$ are replaced by $(a|\nabla|x)$, $(a|\nabla|i)$, and $(x|\nabla|i)$, respectively.

\subsection{Basis set considerations}
\label{sec:basis}

As noted in the introduction, one of the major challenges in calculation of the Bethe logarithm is
the choice of the basis set functions for the expansion of $\Psi(k)$. In this work, we employ the Slater-type orbitals (STOs) basis set defined as
\begin{align}
\label{sto1}
\chi_{nlm}(\textbf{r};\zeta) = r^{n-1} e^{-\zeta r}\,Y_{lm}(\hat{\mathbf{r}}),
\end{align}
where $\zeta>0$, and $Y_{lm}$ are the real spherical harmonics. Normalization constant is not included in the above definition, but normalized orbitals are used in all calculations reported in this paper. For the expansion of the Hartree-Fock orbitals, we use the canonical STOs ($1s$, $2p$, $3d$, etc.) with the principal and angular quantum numbers related by $l=n-1$. However, for the expansion of the response function $\Psi(k)$, we additionally employ STOs with $l=n$, i.e. $1p$, $2d$, and so on. The reasons behind this choice shall be explained shortly, but first we discuss some of their relevant properties. For brevity, we refer to the STOs with $l=n$ as \emph{atypical} STOs.

Let us consider the simplest member of the atypical STOs, namely the $1p$ functions. While $0s$ functions (with $n=l=0$) appear to be the first atypical STO, they do not have a finite kinetic energy, and hence cannot be used as a basis set for quantum chemical calculations. An alternative definition of $0s$ functions was given by Szalewicz and Monkhorst~\cite{szalewicz81}, but it deviates from the standard formula in Eq.~(\ref{sto1}). Written explicitly in the Cartesian coordinates, the $1p$ function takes the form (up to a trivial multiplicative constant):
\begin{align}
\label{sto1p}
 \chi_{1p_\mu}(\textbf{r};\zeta) = \frac{\mu}{r}\,e^{-\zeta r},
\end{align}
where $\mu=x,y,z$. An interesting property of $1p$ functions is that they are not everywhere continuous. For example, for $x=y=0$ the $1p_z$ function reads $e^{-\zeta r}\, \mbox{sgn}(z)$, where $\mbox{sgn}$ stands for the sign function. However, this causes no major problems in 
computation of the necessary matrix elements since discontinuities appear only on subsets of $\mathbb{R}^3$ that are of measure zero. Similar analysis can be performed for atypical STOs with higher principal quantum numbers, i.e. $2d$, $3f$, etc. While they are continuous everywhere, unlike $1p$, the discontinuities appear in their derivatives.

To show how the atypical STOs naturally arise in the present context, we return to the response equation~(\ref{psik}). Difficulties in the calculation of $\Psi(k)$ in a finite basis set are encountered only when $k$ is large. Therefore, let us consider the analytic behavior of the function $\Psi(k)$ as $k\rightarrow\infty$. As shown in Refs.~\cite{schwartz61,bukowski92}, it can be represented in the form
\begin{align}
\label{phikasym}
 \Psi(k) \rightarrow \frac{1}{k} \nabla \Psi_0 + \frac{1}{k^2} U(k),
\end{align}
where $U(k)$ is the so-called Schwartz function. This function is defined as a solution of a  complicated differential equation~\cite{schwartz61,bukowski92} and has no closed-form analytic solution. However, as argued in Ref.~\cite{bukowski92}, it can be accurately approximated by
\begin{align}
\label{uk}
 U_S(k) = \Psi_0 \sum_i \frac{\textbf{r}_i}{r_i^3} \left[ 1 - e^{-\xi r_i}(1+\xi r_i) \right],
\end{align}
with $\xi=\sqrt{2k}$. This function captures the essential analytic features of the exact $U(k)$. Therefore, the basis set used for the expansion of $\Psi(k)$ in Eq. (\ref{psik}) must be able to model both $\nabla \Psi_0$ and $U_S(k)$ effectively.

Let us consider the first term on the right-hand-side of Eq. (\ref{phikasym}). In the simplest approximation, the Hartree-Fock orbitals of many-electron atoms can be approximated by linear combinations of hydrogen-like orbitals with some effective nuclear charge $Z_{\mathrm{eff}}$ (Slater screening constants~\cite{slater30}). For the first- and second-row atoms it is sufficient to consider the $1s$, $2s$, $3s$, $2p$ and $3p$ hydrogen-like states. The gradients of the $ns$ orbitals (in the Cartesian representation) read
\begin{align}
    &\nabla\phi_{1s}(\mathbf{r}) \propto \frac{\mathbf{r}}{r}\,e^{-Z r}, \\
    &\nabla\phi_{2s}(\mathbf{r}) \propto \frac{\mathbf{r}}{r}\,e^{-Z r/2} (Z r - 4), \\
    &\nabla\phi_{3s}(\mathbf{r}) \propto 
    \frac{\mathbf{r}}{r}\,e^{-Z r/3} (2Z^2r^2 - 30Zr + 81),
\end{align}
where the symbol $\propto$ stands for ``proportional to''. Clearly, differentiation of $ns$ orbitals generates a $1p$ orbital plus a sum of non-singular terms which can be described using the ordinary basis. Of course, one can adopt a brute-force approach and attempt to approximate $\nabla\phi_{ns}(\mathbf{r})$ by using the standard $2p_m$ functions only. However, our numerical experiments show that this is inefficient due to the aforementioned lack of continuity.

The gradients of the $np$ wavefunctions are found similarly. Differentiation leads to $2d$ orbitals plus a regular part which can be accurately represented in a canonical STOs basis (provided that the non-linear parameters are optimized properly). To sum up, in order to represent the quantity $\nabla \Psi_0$ efficiently within the STOs basis, both canonical and atypical STOs are required. Of course, if one goes beyond the model of atomic orbitals represented by hydrogen-like wavefunctions, there is no guarantee that just a single $1p$ orbital per occupied $ns$ state will be sufficient to represent the gradient accurately. The number of additional atypical STOs and their exponents must be found by trial and error, and numerical optimization.

The second part of the analysis concerns the Schwartz function, Eq.~(\ref{uk}). When the electron is far from the nuclei this function can be represented in the canonical basis set without major difficulties. Therefore, we need to consider its behavior when the electron-nucleus distance is small. Expanding Eq.~(\ref{uk}) in a series around $r_i=0$ and neglecting the multiplicative constants we obtain
\begin{align}
U_S(k) \approx \Psi_0\sum_i\frac{\mathbf{r}_i}{r_i} e^{-\xi r_i} + \mathcal{O}(r_i),
\end{align}
This leads to the conclusion that in a vicinity of the nuclei, the Schwartz function can be approximated by the ground state wavefunction multiplied by a sum of $1p$ orbitals. This means that the one-electron basis used in the calculations must be able to model products of each occupied Hartree-Fock orbital and a $1p$ function. Taking into account that we consider atoms with occupied orbitals of $1s$, $2s$, $3s$, $2p$, and $3p$ symmetry, this leads to the conclusion that $1p$ and $2d$ orbitals must be included in the basis set for the expansion of the response function. An estimate of the exponents of the $1p$ and $2d$ functions is $\xi=\sqrt{2k}$.

\section{Numerical calculations}
\label{sec:numerics}

\subsection{Computational details}
\label{sec:kompot}

\begin{table}[t]
\caption{\label{tab:hfatoms}
Comparison of the optimized Hartree-Fock energies with the best literature estimates. All values are given in atomic units.}
\begin{ruledtabular}
\begin{tabular}{cll}
 atom & \multicolumn{2}{c}{Hartree-Fock energy} \\
      & \multicolumn{1}{c}{this work} 
      & \multicolumn{1}{c}{literature~\cite{saito09,cinal20}} \\
 \hline\\[-1.2em]
 He & \phantom{00}$-$2.861 679 995 612 232 & \phantom{00}$-$2.861 679 995 612 238 9 \\
 Li & \phantom{00}$-$7.432 726 930 729 724 & \phantom{00}$-$7.432 726 930 73 \\
 Be & \phantom{0}$-$14.573 023 168 311 & \phantom{0}$-$14.573 023 168 316 400 \\
 B  & \phantom{0}$-$24.529 060 728 530 & \phantom{0}$-$24.529 060 728 5 \\
 \hline\\[-1.2em]
 C  & \phantom{0}$-$37.688 618 962 972 & \phantom{0}$-$37.688 618 963 0 \\
 N  & \phantom{0}$-$54.400 934 208 516 & \phantom{0}$-$54.400 934 208 5 \\
 O  & \phantom{0}$-$74.809 398 470 020 & \phantom{0}$-$74.809 398 470 0 \\
 F  & \phantom{0}$-$99.409 349 386 718 & \phantom{0}$-$99.409 349 386 7 \\
 \hline\\[-1.2em]
 Ne & $-$128.547 098 109 381 9 & $-$128.547 098 109 382 042 \\
 Na & $-$161.858 911 616 526 & $-$161.858 911 617 \\
 Mg & $-$199.614 636 423 742 & $-$199.614 636 424 506 710 \\
 Ar & $-$526.817 512 801 437 & $-$526.817 512 802 723 355 \\
\end{tabular}
\end{ruledtabular}
\end{table}

Calculation of one- and two- integrals with Slater-type orbitals has been considered
in Refs.~\cite{lesiuk12,lesiuk14a,lesiuk14b,lesiuk16} and this formalism is adopted in the present work. 
The inclusion of the atypical STOs does not lead to fundamentally new types of matrix elements in the case of atomic systems. The necessary integrals are either evaluated directly in the same way as in Refs.~\cite{lesiuk12,lesiuk14a,lesiuk14b,lesiuk16} or obtained by analytic continuation of the formulas derived for the conventional STOs.

The basis sets used for the expansion of the Hartree-Fock orbitals are constructed out of the canonical STOs. The composition of the basis sets is twelve $1s$ orbitals for helium, fourteen $1s$ orbitals for lithium and beryllium, sixteen $1s$ and $2p$ for B-Ne, and eighteen $1s$ and $2p$ for Na, Mg, Ar. The number of basis set functions was chosen after extensive numerical trials, aiming at the accuracy of the total energy of around 12 significant digits.
The exponents $\zeta_{kl}$ of the STOs with the angular momentum $l$ were restricted to form a geometric sequence, i.e.~$\zeta_{kl}=\alpha_l\beta_l^k$. The parameters $\alpha_l$ and $\beta_l$ were fully optimized employing the variational principle. The total Hartree-Fock energies obtained by us are given in Table~\ref{tab:hfatoms}, and are compared with the best literature references. In some cases, the energies reported here appear to be the most accurate to date. The computer code used for all calculations reported in this work is available online~\cite{github} on the \textsc{GitHub} platform.

By solving the Hartree-Fock equations, one obtains a set of virtual orbitals as a byproduct. As explained in Sec.~\ref{sec:response}, they are used to form excited-state determinants into which the response function is expanded. However, at this stage, the virtual basis has to be augmented with additional STOs necessary for accurate representation of the response function. Unfortunately, as the functions added to the basis are, in general, neither orthogonal to the occupied orbitals nor orthonormal, the response equations derived in Sec.~\ref{sec:response} would no longer valid in the extended basis. To eliminate this problem, an intermediate preparatory stage, comprising three steps, is introduced before the response calculations are initiated. First, the occupied orbital space is projected out from each additional function by applying the operator $\hat{O}=1-\sum_i |i\rangle\langle i|-\sum_x |x\rangle\langle x|$.  Next, the projected functions are added to the virtual space which is then orthonormalized using the L\"owdin algorithm~\cite{lowdin50}. Lastly, the virtual-virtual block of the Fock matrix is evaluated, see Eq.~(\ref{rcoupling}), and diagonalized to form the final virtual basis. 

There is a slight disadvantage of the proposed procedure which must be mentioned. While in the new basis the occupied-occupied and virtual-virtual blocks of the effective Fock operator are diagonal, the closed-virtual and open-virtual blocks are no longer strictly zero. However, the magnitude of the off-diagonal terms decreases rapidly as the quality of the basis used for the expansion of the occupied orbitals is improved, and vanishes in the complete basis set limit. For the basis sets used in this work, the absolute magnitude of the off-diagonal terms in relation to the corresponding diagonal elements never exceeded $10^{-6}$. Therefore, the spurious off-diagonal elements were neglected in all subsequent calculations.

The basis set used for the expansion of the response functions is created by adding two groups of functions. First, the expansion of the response functions requires a basis with angular momentum higher by one unit than the occupied orbitals. To account for that, we take all functions with the highest angular momentum present in the Hartree-Fock basis and replicate them increasing their $l$ by one. Second, the atypical STOs must be added to the basis. We found that to obtain accurate results, it is sufficient to add two $1p$ functions per each occupied $ns$ shell in the atom, and two $2d$ per each occupied $np$ shell. For example, for lithium or beryllium atoms four $1p$ functions are added, while for sodium or magnesium - six $1p$ functions and two $2d$ functions. The exponents of the atypical STOs are fully optimized for each $k$ separately. The remaining functions in the basis are independent of $k$ and are not affected by the optimization. This approach is a middle ground between the full optimization of the basis for each $k$, which is prohibitively computationally costly, and using a fixed basis for each $k$ which does not offer sufficient flexibility.

The working formula for calculating the Bethe logarithm, Eq.~(\ref{lnk0_t}), involves integration over the variable $t$ which must be carried out numerically. Our preliminary tests showed that the standard Gauss-Legendre quadrature (with a linear transformation of the integration interval) with fifty nodes is capable of providing relative accuracy of at least six significant figures. This setup is used in all calculations presented in this work.

\subsection{Benchmark calculations: hydrogen atom}
\label{sec:bench-hydrogen}

Before considering many-electron atoms, the performance of the proposed formalism was tested in calculations for systems where reliable reference values are known. In this and next section we provide results for the hydrogen and helium atoms, and compare them with the data available in the literature.

The hydrogen atom is an ideal example for testing the adequacy of the basis set used for the expansion of the response function. Indeed, the exact ground-state wavefunction is known in this case and can be used in the calculations without any difficulties. This makes the accuracy of the results dependent
solely on the quality of the basis used for the expansion of the response function, eliminating other sources of error. The basis set used for this expansion comprises twenty $2p$ STOs with exponents being a geometric progression optimized to fit the first $30$ excited states of the hydrogen atom of the $p$ symmetry. Additionally, the basis includes two $1p$ functions with exponents optimized for each $k$. 

\begin{figure}[t!]
 \hspace{-0.5cm}
 \includegraphics[scale=0.65]{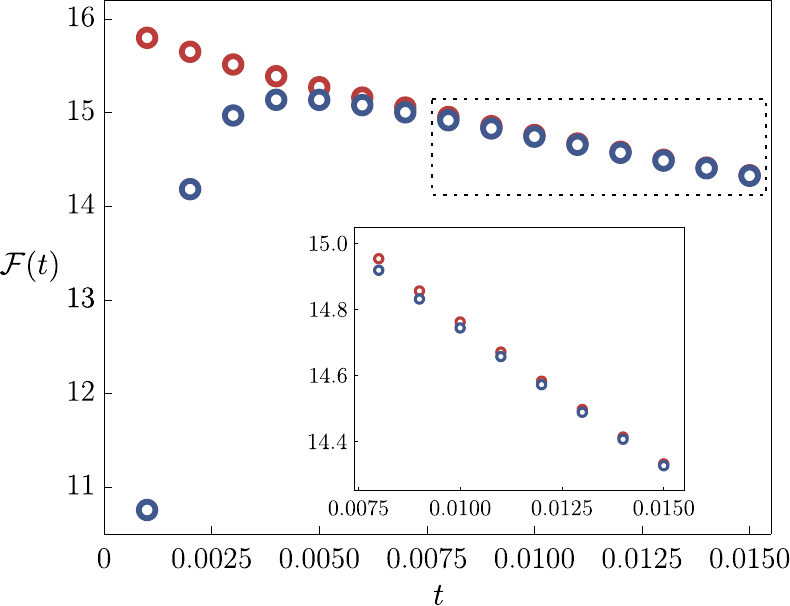}
 \caption{\label{fig:hydrogenft} Values of the integrand $\mathcal{F}(t)$ in Eq.~(\ref{lnk0_t}) obtained with a basis set including one $1p$ function (orange dots) and two $1p$ functions (red dots) as a function of the parameter $t$. The inset graph presents a magnification of the plot within range of approximately $t\in(0.0075,0.0150)$. For larger $t$, the two data sets are indistinguishable on the scale of the plot.}
\end{figure}

In the numerical integration of Eq.~(\ref{lnk0_t}) one requires to calculate the integrand $\mathcal{F}(t)$ at grid points where the value of $t$ is close to zero. For small $t$, it is impractical to evaluate $\mathcal{F}(t)$ directly from Eqs.~(\ref{lnk0_t})~and~(\ref{ft}), because the former formula involves a substantial cancellation between the terms in the numerator. To circumvent this problem we construct a fit of the function $\mathcal{F}(t)$ in a region where $t$ is moderately small, and calculate $\mathcal{F}(t)$ for smaller $t$ by extrapolation of the fitting function instead of the direct evaluation.

To this end, in the first stage of the calculations we compute values of $\mathcal{F}(t)$ in a region where $t$ is small, so that the series expansion~(\ref{ftsml}) valid, but large enough to avoid the aforementioned cancellation problem. For the hydrogen atom we used $26$ equidistant points between $t=0.005$ and $t=0.030$. These values are then fitted with the formula
\begin{align}
\label{Ftfit}
 \mathcal{F}(t) = f_3 + f_4 \,t \ln t + f_5\, t + f_6\, t^2 \ln t + \ldots
\end{align}
which follows from Eqs.~(\ref{lnk0_t})~and~(\ref{ftsml}) by cancelling the counter-terms. With seven terms, i.e. up to and including $t^3$, the largest absolute deviation of the fit from the raw data points is roughly $6\cdot10^{-10}$, with the mean absolute deviation of only about $2\cdot10^{-10}$. The final results were virtually unaffected when five or nine terms were used in Eq.~(\ref{Ftfit}). It is also worth pointing out that for the hydrogen atom the exact values of the first two coefficients in Eq. (\ref{Ftfit}) are known, namely $f_3=16$ and $f_4=32$. From the fitting function, we obtained $f_3=16.0000025$ and $f_4=32.0024$, in a remarkably good agreement with the exact results. This confirms the accuracy of Eq.~(\ref{Ftfit}) and proves that the problem of small $t$ can be avoided by fitting.

In the second stage of the calculations, the remaining $f(t)$ and $\mathcal{F}(t)$ required 
by the quadrature are computed. For $t>0.030$, i.e., outside of the region covered by the fit, the 
conventional expressions~(\ref{lnk0_t})~and~(\ref{ft}) are used as they stand. In this way, we 
obtain the final value of $\ln k_0=2.98\,412\,794$ for the hydrogen atom 
in the ground state. This is compared with the reference value obtained by Drake and 
Goldman~\cite{drake00}, $\ln k_0=2.98\,412\,856$ (all digits shown are accurate), revealing that our result is accurate to six significant digits (the relative error is about 0.2 ppm). This accuracy level is more than sufficient from the point of view of calculations for many-electron systems.

It is interesting to study the influence of the atypical STOs on the quality of the results obtained for the hydrogen atom. To this end, we performed calculations of the integrand $\mathcal{F}(t)$ using a basis set without any $1p$ functions, and, separately, with only one $1p$ function. When no $1p$ functions are included in the basis, the results are clearly nonsensical and diverge to a large negative value for small $t$. Therefore, they are not considered here. When only one $1p$ function is used, the results for $t\approx 0.020$ are almost identical to the data discussed previously, as shown in Fig.~\ref{fig:hydrogenft}. For example, around $t=0.020$ they differ by only about several parts in ten thousands. However, as the value of $t$ gets closer to zero, the results obtained with only one $1p$ function in the basis deteriorate. This is a typical symptom of a basis set incompleteness problem in the computation of 
$\mathcal{F}(t)$. Instead of converging to the correct limit, $\mathcal{F}(0)=16$, a rapid divergence a towards large negative value is observed. No such artifacts are present in calculations employing the basis set with two $1p$ functions. One may also ask to what degree adding further $1p$ functions improves the accuracy in comparison with the data shown in Fig.~\ref{fig:hydrogenft}. We found that the addition of the third $1p$ function is not worth the effort related to the optimization of the exponents. The differences are too small to be visible on the scale of the plot in Fig.~\ref{fig:hydrogenft}, but the accuracy of the $\ln k_0$ is improved by one significant digit.

\subsection{Benchmark calculations: helium atom}
\label{sec:bench-helium}

\begin{table}[t]
\caption{\label{tab:helium}
Results of the calculations for the helium atom with the electronic wavefunction approximated by the Hartree-Fock determinant (see Sec.~\ref{sec:bench-helium} for details of the procedure). All values are given in atomic units.}
\begin{ruledtabular}
\begin{tabular}{cl}
 quantity & \multicolumn{1}{c}{value} \\
 \hline\\[-2.2ex]
 $-E_0$                                  & \phantom{0}$2.86\,167\,999\,561$ \\
 $-\langle\Psi_0|\nabla^2|\Psi_0\rangle$ & \phantom{0}$5.72\,335\,999\,122$ \\
 $\mathcal{D}$ -- formula~(\ref{exd})    & $45.18\,764\,401\,403$ \\
 $\mathcal{D}$ -- formula~(\ref{newden}) & $45.18\,763\,551\,604$ \\
 $\ln k_0$                               & \phantom{0}$4.39\,124$ \\
\end{tabular}
\end{ruledtabular}
\end{table}

The next important test case is the helium atom. The overall procedure for calculation of $\ln k_0$ 
is analogous as for the hydrogen atom. The major differences stem from the fact that for helium the Hartee-Fock wavefunction, i.e. a single Slater determinant, is not the exact solution of the electronic Schr\"odinger equation. Because of that, the denominator in Eq.~(\ref{bethe0}) cannot be calculated using the exact formula~(\ref{exd}). Instead, the approach based on the resolution of identity (RI) described in Sec.~\ref{sec:schwartz} is employed. In Table~\ref{tab:helium} we report values of the denominator calculated using both approaches, as well as several other key quantities required in the computation of the Bethe logarithm. A remarkable agreement is obtained between values of $\mathcal{D}$ obtained according to both formulas. This suggests that for helium, Eq.~(\ref{exd}) is sufficiently accurate even if an approximate electronic wavefunction is used. However, we found that this conclusion is no longer true in calculations for many-electron atoms reported further. As the evaluation of the denominator according to Eq.~(\ref{newden}) is not computationally demanding, we continue to use it in the remainder of this work.

Our final result for the Bethe logarithm of the helium atom is $\ln k_0(\mbox{He})=4.39\,124$. It agrees to about $0.5\%$ with the reference value of Korobov~\cite{korobov19}, namely $\ln k_0(\mbox{He})=4.370\,160$ (all digits shown are exact). The main source of error is the accuracy of the ground-state electronic wavefunction represented by the Hartee-Fock determinant. Nonetheless, this level of accuracy is sufficient in many calculations for larger systems.

The fact that the Bethe logarithm can be calculated with accuracy better than 1\% using the Hartee-Fock method may be surprising at first. For example, when the Hartee-Fock method is used for calculation of similar quantities involving a response theory, such as static or frequency-dependent polarizabilities, errors are typically by at least an order of magnitude larger~\cite{lesiuk20b,lesiuk23}. The fact that the electron correlation effects, not accounted for in the Hartree-Fock method, are of minor importance in calculation of the Bethe logarithm, can be understood by noting that the response function $\Psi(k)$ is concentrated primarily around the atomic nucleus, especially for large $k$. This regime is dominated by the electron-nuclear cusp resulting from the Coulomb singularity of the electron-nucleus interaction potential, not by the electronic cusp that requires a correlated treatment beyond the mean-field approximation.

\subsection{Many-electron atoms}
\label{sec:many-atoms}

\begin{table}[t]
\caption{\label{tab:many-atoms}
Results of the calculations for many-electron atoms. All values are given in atomic units.}
\begin{ruledtabular}
\begin{tabular}{crrc}
 atom & \multicolumn{3}{c}{quantity} \\
 & \multicolumn{1}{c}{$-\langle\Psi_0|\nabla^2|\Psi_0\rangle$}
 & \multicolumn{1}{c}{$\mathcal{D}$}
 & $\ln k_0$ \\
 \hline\\[-2.2ex]
 Li & $14.865\,454$  & $260.403$      & $5.194$ \\
 Be & $29.146\,046$  & $889.390$      & $5.763$ \\
 B  & $48.248\,405$  & $2\,257.521$   & $6.339$ \\
 C  & $72.588\,886$  & $4\,805.053$   & $6.706$ \\
 N  & $102.443\,892$ & $9\,056.788$   & $6.973$ \\
 O  & $137.919\,402$ & $15\,665.784$  & $7.220$ \\
 F  & $179.393\,816$ & $25\,352.026$  & $7.415$ \\
 Ne & $227.138\,262$ & $38\,994.224$  & $7.581$ \\
 Na & $282.111\,532$ & $57\,691.053$  & $7.770$ \\
 Mg & $344.007\,915$ & $82\,558.373$  & $7.943$ \\
 Ar & $861.417\,446$ & $434\,267.898$ & $8.761$ \\
\end{tabular}
\end{ruledtabular}
\end{table}

% A natural perpetuation of applications of our method for the Bethe logarithm computation are calculations of many-electron atoms.
Next, we pass to the calculation of the Bethe logarithm for many-electron atoms. 
In these calculations we closely follow the protocol established and verified in the preceding sections for hydrogen and helium. In Table~\ref{tab:many-atoms} we report the obtained values of $\ln k_0$ and of several key quantities encountered in the formalism.

In the case of lithium and beryllium atoms, accurate values of the Bethe logarithm were obtained by Pachucki and Komasa~\cite{pachucki03,pachucki04}, namely $\ln k_0(\mbox{Li})=5.178\,17(3)$ and $\ln k_0(\mbox{Li})=5.750\,34(3)$, using explicitly correlated wavefunction models. It is worth pointing out that our results deviate by only $0.3\%$ and $0.2\%$, respectively, from this reference. These relative errors are noticeably smaller than in the case of helium discussed in the previous section (ca. $0.5\%$). This suggests that the accuracy of the mean-field approach improves with increasing number of electrons in the atom. This is not entirely unexpected, as in the $Z\rightarrow\infty$ limit, the atomic Bethe logarithm tends to a weighted mean of the hydrogenic values for the corresponding occupied states. The latter are represented very accurately in our formalism, see Sec.~\ref{sec:bench-hydrogen}. Unfortunately, as sufficiently reliable reference data is not available for atoms beyond beryllium, it is impossible to confirm this trend beyond a reasonable doubt.

No significant problems are encountered in the calculations for the remaining atoms. It is difficult to rigorously estimate the error of the calculated values, but the experience gained from benchmark calculations suggests that the accuracy of the data given in Table~\ref{tab:many-atoms} is better than $0.5\%$. It is also interesting to compare our results with a frequently-used approximation which assumes that the dominant contribution to the Bethe logarithm comes from the $1s^2$ shell which is doubly occupied in all atoms. It is hence reasonable to approximate $\ln k_0$ for a many-electron atom by the corresponding value for the helium-like ion with the same nuclear charge. The latter quantity is known to very high accuracy~\cite{goldman83,goldman84,bhatia98,baker08}. A comparison reveals that the agreement for lithium and beryllium is remarkably good, with the difference between the helium-like estimate and the data given in Table~\ref{tab:many-atoms} being a fraction of a percent. However, as the nuclear charge increases, the deviation becomes larger, reaching about $2-3\%$ for carbon, nitrogen and oxygen, and slowly decreasing beyond this point. This error is clearly much larger than of the mean-field approach.

\subsection{Lamb shift in nitrogen molecule}
\label{sec:n2-example}

As an illustrative application of the theory reported in this work, we consider the Lamb shift contribution to the vibrational energy levels of the nitrogen molecule, N$_2$. The knowledge of accurate properties of the nitrogen molecule, the major component of air, is important in various research fields. For example, development of accurate pressure standards based on refractivity or capacitance measurements is an ongoing effort with applications in science and industry alike~\cite{gaiser15,rourke19,gaiser20,gaiser22}. The nitrogen is an ideal candidate for the working gas in such measurements as it is safe, inexpensive and broadly available.

In contrast to atoms, the fundamental properties of molecules, such as polarizability or magnetic susceptibility, are temperature-dependent. This dependence is non-negligible and has to be taken into account in order to be useful from the experimental point of view. In order to model this effect, the properties calculated as a function of the molecular geometry have to be averaged over the accessible vibrational levels. The weight of each level is equal to the Boltzmann factor evaluated at a temperature of interest. From the point of view of the aforementioned applications, it is sufficient to consider temperatures of up to $3000\,$K. Based on the data available in the literature~\cite{leroy06}, one can estimate that the first ten vibrational levels must be taken into account in order to compute the temperature dependence of an arbitrary quantity with relative accuracy of one part per ten thousand.

In order to evaluate the Lamb shifts of these vibrational levels, we developed a potential energy curve for the nitrogen molecule within the Born-Oppenheimer (BO) approximations. The calculations were performed at the internally-contracted multireference configuration interaction (icMRCI) level of theory~\cite{werner82,werner88,knowles88} using doubly-augmented d-aug-cc-pV$X$Z, $X=2,\ldots,6$, family of Gaussian basis sets~\cite{dunning89,kendall92,woon94,wilson96}. For improved accuracy, the basis sets were fully uncontracted and the Davidson cluster correction (MRCI+Q) with relaxed reference was applied~\cite{davidson77,knowles91,stark96}. The leading-order relativistic corrections were taken into account perturbatively using the Cowan-Griffin Hamiltonian~\cite{cowan76}. All calculations were performed using the \textsc{Molpro} program package~\cite{werner12,werner20}. The results were extrapolated to the complete basis set limit using the Riemann extrapolation formula~\cite{lesiuk19b}. 

The leading-order QED correction to the interaction potential was evaluated using the formula
\begin{align}
\label{qed1}
\begin{split}
E_{\mathrm{QED}} &= \frac{8\alpha}{3\pi}\left(\frac{19}{30}-2\ln \alpha - \ln k_0\right)\langle D_1 
\rangle,
\end{split}
\end{align}
which is obtained from Eq.~(\ref{qed}) by dropping all two-electron terms. The molecular Bethe logarithm was approximated by the atomic value reported in Sec.~\ref{sec:many-atoms} according to the discussion given in Sec.~\ref{sec:molbethe}. To illustrate the accuracy of the complete \emph{ab initio} potential, we note the obtained the well depth, $D_e=79\,898.6\,$cm$^{-1}$, and the equilibrium internuclear distance, $R_e=2.073\,$a.u., which are in a good agreement with the experimental results, $D_e=79\,845(9)\,$cm$^{-1}$ and $R_e=2.074\,312(2)~$a.u., taken from Ref.~\cite{roncin84,leroy06}.

\begin{table}[t]
\caption{\label{tab:n2-levels}
Vibrational energies (measured with respect to the bottom of the potential well) and the corresponding Lamb shifts of the first ten vibrational levels of the nitrogen molecule. All values are given in cm$^{-1}$.}
\begin{ruledtabular}
\begin{tabular}{ccc}
 $\nu$ & $E_\nu$ & $\delta_\nu^{\mathrm{QED}}$ \\
 \hline\\[-2.2ex]
0 &  1\,173.3 & 3.3\\
1 &  3\,507.8 & 3.5\\
2 &  5\,813.2 & 3.2\\
3 &  8\,080.7 & 3.1\\
4 & 10\,327.5 & 3.0\\
5 & 12\,549.3 & 3.0\\
6 & 14\,757.1 & 2.8\\
7 & 16\,930.1 & 2.8\\
8 & 19\,061.2 & 2.9\\
9 & 21\,168.1 & 2.6\\
\end{tabular}
\end{ruledtabular}
\end{table}
The vibrational levels were obtained by solving the one-dimensional radial Schr\"odinger equation for the nuclear motion. The standard Numerov method~\cite{cooley1961improved} was applied for this purpose using the cubic spline interpolation of the interaction potential as programmed in LEVEL code\cite{leroy2017level}. We consider only the most abundant isotope of nitrogen, $^{14}$N, with the atomic mass $m(^{14}\mathrm{N})=14.003\,074$, according to Ref.~\onlinecite{wang21}.

In order to find the Lamb shifts of the vibrational levels, denoted $\delta_\nu^{\mathrm{QED}}$ further in the text, the nuclear Schr\"odinger equation was solved with and without the QED correction to the potential, Eq.~(\ref{qed1}). The difference between the corresponding vibrational energies in the two calculations was taken as $\delta_\nu^{\mathrm{QED}}$. In Table~\ref{tab:n2-levels} we report energies of the vibrational levels ($J=0$) obtained with the complete interaction potential, i.e. including the BO, relativistic, and QED components, and the corresponding $\delta_\nu^{\mathrm{QED}}$ shifts for $\nu=0,1,\ldots,9$. Overall, the Lamb shift in nitrogen is of the order of a few cm$^{-1}$ for these states, showing that it is spectroscopically significant and at least an order of magnitude larger than the experimental uncertainties involved in measuring the vibrational transition energies in N$_2$ (see Refs.~\cite{leroy06,roncin99} for a discussion of the available experimental data). Of course, at present the uncertainty of theoretical predictions for N$_2$ is limited by errors in the BO component of the potential, not by the lack of the $E_{\mathrm{QED}}$ component. Nonetheless, the work on improving the BO potential in ongoing.

\section{Conclusions}
\label{sec:conclusions}

In this work we have developed and implemented a method for calculation of the Bethe logarithm for many-electron atoms. The proposed formalism is based on the mean-field representation of the ground-state electronic wavefunction and of the response functions required in the Schwartz method. We have discussed difficulties encountered in the calculations of the Bethe logarithm, in particular those related to the specific basis set requirements in the vicinity of the atomic nucleus. This problem has been ameliorated by introducing a modified basis set of exponential functions which are able to accurately represent the gradient of hydrogen-like orbitals.

We have computed the Bethe logarithm for ground electronic states of atoms from hydrogen to magnesium and, additionally, for argon. The results have been compared with the available reference data from the literature, whenever possible. In general, the mean-field approximation introduces a surprisingly small error in the calculated values, suggesting that the electron correlation effects are of minor importance in determination of the Bethe logarithm. Therefore, one may expect that even a low-level correlated many-body theory may be entirely sufficient to calculate $\ln k_0$ with high accuracy. We believe this is a promising direction of further study.

Finally, we have studied a robust and computationally inexpensive scheme to evaluate the Lamb shift for arbitrary light molecular systems at little computational cost. In this method, a weak dependence of the Bethe logarithm on the molecular geometry is exploited. As an illustration, we have calculated Lamb shifts of the vibrational levels of the nitrogen molecule. The results prove that the QED effects in the spectrum of N$_2$ molecule are spectroscopically significant within the current experimental uncertainties.

\begin{acknowledgments}
The authors are grateful to prof. B. Jeziorski for fruitful discussions and for reading and commenting on the manuscript prior to submission. The project (22IEM04 MQB-Pascal) has received funding from the European Partnership on Metrology, co-financed from the European Union’s Horizon Europe Research and Innovation Programme and by the Participating States. We gratefully acknowledge Poland's high-performance Infrastructure PLGrid (HPC Centers: ACK Cyfronet AGH, PCSS, CI TASK, WCSS) for providing computer facilities and support within computational grant PLG/2022/015748.
\end{acknowledgments}

\bibliography{atomic-bethe}

\end{document}